\documentclass[12pt]{spieman}  
\usepackage{amsmath,amsfonts,amssymb}
\usepackage{graphicx}
\usepackage{setspace}
\usepackage{tocloft}
\usepackage{graphicx}  
\usepackage{dcolumn}  
\usepackage{bm}  
\usepackage{xfrac} 
\usepackage{mathtools} 
\usepackage{euscript} 
\usepackage[range-phrase = \text{ to },  
range-units=single,  
separate-uncertainty = true, 
group-separator = {,},  
group-digits=integer,  
list-units = single  
]{siunitx} 
\usepackage[dvipsnames]{xcolor} 
\usepackage{ctable}  
\usepackage{multirow}  
\usepackage[noabbrev]{cleveref}  

\creflabelformat{equation}{#2\textup{#1}#3}  
\newcolumntype{C}{>{\centering\arraybackslash}X}  
\newcommand*{\sZ}{\ensuremath{\EuScript Z}}
\newcommand*{\sI}{\ensuremath{\mathcal I}}
\newcommand*{\sQ}{\ensuremath{\mathcal Q}}
\DeclarePairedDelimiter{\abs}{\lvert}{\rvert}
\DeclarePairedDelimiter{\brackets}{[}{]}
\DeclarePairedDelimiter{\braces}{\{}{\}}
\DeclarePairedDelimiter{\parens}{(}{)}

\newcommand*{\mat}[1]{\ensuremath{\boldsymbol{\mathrm{#1}}}} 
\renewcommand*{\vec}[1]{\ensuremath{\boldsymbol{\mathit{#1}}}} 
\def\transpose{{\ensuremath{\mathsf{T}}}}

\title{Improving the dynamic range of single photon counting kinetic inductance detectors}

\author[a,*]{Nicholas Zobrist}
\author[b]{Nikita Klimovich}
\author[b]{Byeong Ho Eom}
\author[a]{Gr\'egoire Coiffard}
\author[a]{Miguel Daal}
\author[a]{Noah Swimmer}
\author[a]{Sarah Steiger}
\author[b]{Bruce Bumble}
\author[b]{Henry G. LeDuc}
\author[b]{Peter Day}
\author[a]{Benjamin A. Mazin}

\affil[a]{Department of Physics, University of California, Santa Barbara, CA 93106, USA}
\affil[b]{Jet Propulsion Laboratory, California Institute of Technology, Pasadena, California 91125, USA}

\cftpagenumbersoff{figure}
\cftpagenumbersoff{table} 
\begin{document} 
\maketitle

\begin{abstract}
We develop a simple coordinate transformation which can be employed to compensate for the nonlinearity introduced by a Microwave Kinetic Inductance Detector's (MKID) homodyne readout scheme. This coordinate system is compared to the canonically used polar coordinates and is shown to improve the performance of the filtering method often used to estimate a photon's energy. For a detector where the coordinate nonlinearity is primarily responsible for limiting its resolving power, this technique leads to increased dynamic range, which we show by applying the transformation to data from a hafnium MKID designed to be sensitive to photons with wavelengths in the \SIrange{800}{1300}{nm} range. The new coordinates allow the detector to resolve photons with wavelengths down to \SI{400}{nm}, raising the resolving power at that wavelength from \numrange{6.8}{17}.
\end{abstract}


{\noindent \footnotesize\textbf{*}\linkable{nzobrist@physics.ucsb.edu} }

\section{Introduction} \label{sec:intro}
An MKID is a superconducting resonator sensitive to incident radiation through perturbations in its surface impedance.~\cite{Zmuidzinas2012} Broken Cooper pairs inside the superconductor increase the resonator's kinetic inductance and microwave loss, resulting in changes to its resonance frequency, $f_r$, and internal quality factor, $Q_i$. When the signal is a constant flux of photons, the fractional change of the resonance frequency, $\sfrac{\delta f_r}{f_r}$, and the change of the inverse internal quality factor, $\delta Q_i^{-1}$, are proportional to the absorbed power in the detector. However, if an MKID's responsivity is high enough, $\sfrac{\delta f_r}{f_r}$ and $\delta Q_i^{-1}$ resolve the individual photon absorption event. A photon's energy, then, can be extracted from the overall size of the signal. 
These types of MKIDs are called single photon counting, and examples of them include lumped element optical to near-IR detectors,~\cite{Zobrist2019a, Walter2018, Meeker2018, Szypryt2017b, Mazin2013} X-ray thermal kinetic inductance detectors,~\cite{Yan2016, Ulbricht2015, Miceli2014, Quaranta2013} and position dependent strip detectors.~\cite{Mazin2008a, Mazin2006}

The properties of an MKID are encoded in the forward scattering matrix element, $S_{21}$, of the resonator. We extract these properties by fitting $S_{21}$ as a function of the probe frequency, $f_g$, from a microwave signal generator.
\begin{equation} \label{eq:S21}
    S_{21}(f_g) = \frac{Q_c + 2 i Q_c Q_i (x + x_a)}{Q_c + Q_i + 2 i Q_c Q_i x}
\end{equation}
$Q_c$ is the coupling quality factor, related to the strength of the interaction with the transmission line, and $x = \sfrac{(f_g - f_r)}{f_r}$ is the fractional detuning of the generator frequency. $x_a$ parameterizes the resonance asymmetry caused by impedance mismatches and stray couplings in the system.~\cite{Geerlings2013, Khalil2012} This functional form for $S_{21}$ ensures that $x_a$ is independent of changes in the total quality factor, $Q = (Q_c^{-1} + Q_i^{-1})^{-1}$, which will be important as we discuss the signal coordinates.

Nonlinearities associated with the energy circulating inside of the resonator can modify the forward scattering matrix element. For the resonators discussed later in this paper, a reactive nonlinearity must be included which sets up a feedback effect between generator and resonance frequencies.~\cite{Swenson2013} The generator detuning becomes a function of the microwave power in the resonator, modifying our previous expression for $x$ to
\begin{equation} \label{eq:x_nonlinear}
    x = \frac{f_g - f_{r, 0}}{f_{r, 0}} + \frac{a / Q}{1 + 4 Q^2 x^2}.
\end{equation}
Here, $a$ is a unitless parameter proportional to the power of the generator tone and $f_{r, 0}$ represents the low power resonance frequency. 

Tracking the signals $\sfrac{\delta f_r}{f_r}$ and $\delta Q_i^{-1}$ in real-time requires complex algorithms and hardware. Moreover, these systems typically do not have the required bandwidth to readout single photon detectors.~\cite{Kernasovskiy2018} Instead, a homodyne readout scheme is typically used to downconvert the detector response to a finite bandwidth around each readout tone, resulting in an in-phase, \sI, and quadrature, \sQ, signal from a mixer. In the adiabatic case, where changes to \sI\ and \sQ\ occur at frequencies $f \ll \sfrac{f_r}{(2 Q)}$, and after calibrating out effects from amplification and cabling, $\sZ = \sI + i \sQ$ is proportional to the forward scattering matrix element of the resonator.~\cite{Thomas2015}

\begin{figure}[t]
    \includegraphics[width=\linewidth]{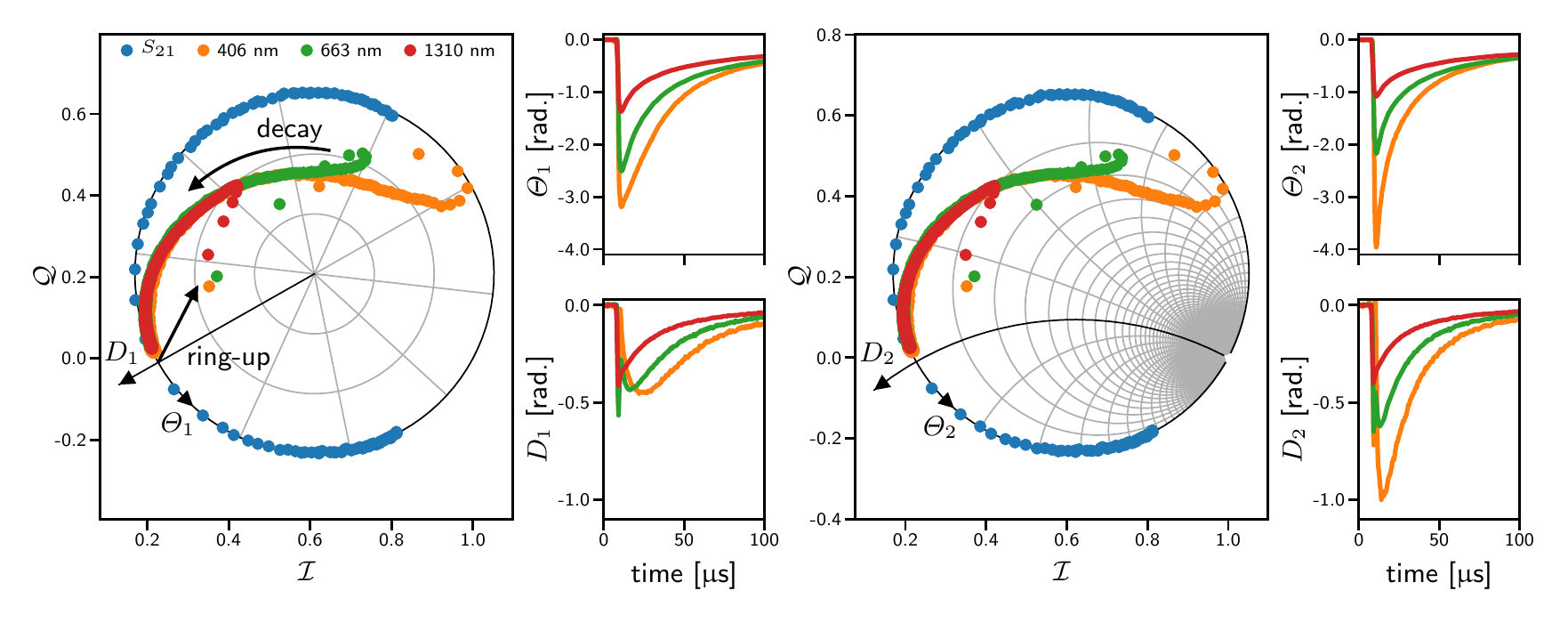}
    \caption{Two coordinate systems are represented with respect to a hafnium MKID, described in detail in reference \citenum{Zobrist2019a}. Averaged response trajectories for three different photon energies are plotted in the \sI\ / \sQ\ plane. Their dependence on time are also shown to the right of each main plot. \textit{Left:} The $\varTheta_1$ and $D_1$ coordinates do not appropriately decompose the larger signal into components that scale with the photon energy. \textit{Right:} Both the compression in phase and dissipation are corrected for with the $\varTheta_2$ and $D_2$ coordinates introduced in \cref{sec:coordinates}, leaving only nonlinearities intrinsic to the detector itself like the energy dependent decay time in $\varTheta_2$ and the delayed response in $D_2$. These coordinates are normalized such that they are approximately equal to the $\varTheta_1$ and $D_1$ coordinates near the resonator's equilibrium point. Their units are written as radians to identify this scaling.}
    \label{fig:coordinates}
\end{figure}

When analyzing photon events, \sI(t) and \sQ(t) are often transformed into a polar coordinate system, referenced to the loop center and radius. Mathematically, they are represented by the following equations:
\begin{subequations}
\begin{align}
    \varTheta_1(t) &\equiv \arg \brackets*{\frac{1 - \sZ(t) - \frac{Q}{2 Q_c} + i Q x_a}{1 - S_{21}(f_g) - \frac{Q}{2 Q_c} + i Q x_a}}  \label{eq:Theta1} \\
    D_1(t) &\equiv \frac{\abs*{1 - \sZ(t) - \frac{Q}{2 Q_c} + i Q x_a}}{\abs*{\frac{Q}{2 Q_c} - i Q x_a}} - 1. \label{eq:D1}
\end{align}
\end{subequations}
\Cref{fig:coordinates} (left) shows these coordinates with respect to $S_{21}$ and an example photon response trajectory. 

$\varTheta_1$ and $D_1$ are commonly referred to as phase and dissipation coordinates since in the adiabatic limit they are proportional to the fractional detuning and dissipation perturbations. We can see this by setting $Z(t) = S_{21}$ with $x \rightarrow x + \delta x(t)$ and $Q_i^{-1} \rightarrow Q_i^{-1} + \delta Q_i^{-1}(t)$ in \cref{eq:Theta1,eq:D1}. To first order in the small signals this approximation gives
\begin{subequations}
\begin{align}
    \varTheta_1(t) &\approx \frac{-4 Q}{1 + 4 Q^2 x^2} \; \delta x(t) \label{eq:dTheta1}\\
    D_1(t) &\approx \frac{-2 Q}{1 + 4 Q^2 x^2} \; \delta Q_i^{-1}(t). \label{eq:dD1}
\end{align}
\end{subequations}
Note that the parameters $x$ and $Q$ are the fit parameters from the original fit to $S_{21}$ and are constant. For larger signals, however, \cref{eq:Theta1,eq:D1} are not proportional to $\delta x$ and $\delta Q_i^{-1}$. The signals mix and saturate as the resonance frequency moves further away from $f_g$. Even worse, $D_1$ is not monotonic as a function of $\delta Q_i^{-1}$.

In the next sections, we explore how these effects can degrade a detector's resolving power---defined as the photon energy over the full-width half-max spread in the estimated energy. The paper is organized as follows: \Cref{sec:energy_estimation} discusses the derivation of the filtering method for estimating the photon energy and why it is often preferred over more complicated analysis methods. In \cref{sec:coordinates}, we address some of the issues with using the polar coordinates with this filtering method by introducing a new data coordinate system. Finally, these new coordinates are applied to data from a hafnium optical to near-IR MKID in \cref{sec:data}, and the resulting resolving power is compared to that obtained with $\varTheta_1$ and $D_1$.

\section{Photon Energy Estimation} \label{sec:energy_estimation}
In principle, the nonideal behavior of the coordinates can be accounted for by properly modeling the detector response. The most general model with a one-to-one correspondence to the photon energy, $E$, may be written as
\begin{equation} \label{eq:model}
    \vec{m}(t; \vec{\xi}) = \vec{r}(t - t_0; E) + \vec{\delta},
\end{equation}
where $t_0$ is the photon arrival time and $\vec{r}$ is a two-component vector containing the energy dependent phase and dissipation response. Since we can only measure data in finite time intervals, a constant offset $\vec{\delta}$ is also included as a parameter to model noise at frequencies below the dataset's bandwidth. For brevity, we use $\vec{\xi}$ to denote all of the parameters in the model.

By using \cref{eq:model}, we are ignoring the possibility that there may be many different responses for a given energy. In MKIDs the most common cause of this effect is from a nonuniform detector response or phonon losses, both of which result in an unavoidable degradation of the detector's resolving power.~\cite{Ulbricht2015, Zobrist2019} Since these effects cannot be addressed by any analysis technique we do not consider them further. Systematic errors like gain and response drifts can also invalidate this model. However, since these variations can be corrected for when they exist~\cite{Fowler2016} and are not present in the data discussed in the next section, they are, likewise, not considered further.

Having a response model allows us to construct a maximum likelihood estimate for the absorbed photon energy by assuming Gaussian noise and minimizing $\chi^2$ with respect to $\vec{\xi}$.
\begin{equation} \label{eq:chi2}
    \chi^2(\vec{\xi}) = \int^\infty_{-\infty} \! \! dt \int^\infty_{-\infty} \! \! dt' \vec{\Delta}^\transpose(t; \vec{\xi}) \ \mat{C}^{-1}(t, t'; \vec{\xi}) \ \vec{\Delta}(t', \vec{\xi})
\end{equation}
$\vec{\Delta}(t; \vec{\xi}) = \vec{d}(t) - \vec{m}(t; \vec{\xi})$ represents the residual between the data, $\vec{d}$, and model. $\mat{C}$ is the covariance matrix that, in the case of nonstationary noise, depends on the model parameters including the photon energy.

Under the above assumptions and assuming that all photon events are well isolated in time, this strategy for computing the photon energy results in an unbiased energy estimator with the minimum possible variance even after digitizing to discrete time steps.~\cite{Fowler2015, Alpert2013} However, $\vec{r}(t; \vec{\xi})$ and $\mat{C}(t, t'; \vec{\xi})$ can be challenging to construct as continuous functions of their parameters. Generative physical models,~\cite{Fowler2018, Shank2014} principal component analyses,~\cite{Yan2016, Busch2016} and template interpolation~\cite{Fixsen2004, Fixsen2014} have been suggested as ways to properly formulate these functions. However, minimizing \cref{eq:chi2} then becomes a nonlinear optimization problem which cannot be solved in real-time.

Detector arrays with many pixels and with high count rates require real-time photon energy estimation to reduce the memory required to save a dataset. This condition restricts the analysis to linear algorithms consisting of, for example, filtering operations or matrix multiplications. Tangent filtering has been suggested as one way to linearize a model around a particular energy but also requires several tangent plane calculations to fully cover a detector's bandwidth.~\cite{Fowler2017} 

If the response shape is a constant function of energy, though, minimizing \cref{eq:chi2} simplifies to a more robust linear filtering method for determining the photon energy. This is the case for the linear photon response model is given by
\begin{equation} \label{eq:linear_model}
    \vec{m}(t; \vec{\xi}) = A(E) \ \vec{s}(t - t_0) + \vec{\delta},
\end{equation}
where $\vec{s}$ is the template response shape normalized so that the calibration function, $A(E)$, represents the sum of the response amplitudes in each dimension for a photon with energy $E$. $A(E)$ converts between energy and detector response and is typically computed by averaging many photon events together at different energies and interpolating between them. $\vec{s}$ is measured similarly but is independent of $E$. Assuming stationary noise, $\vec{s}(t)$ is used to make the filter function, $\vec{h}(t)$.
\begin{equation} \label{eq:filter}
    \vec{\tilde h}(f) = \begin{cases} \frac{\mat{J}^{-1}(f) \ \vec{\tilde s}(f)}{\int_{f \neq 0} df \ \vec{\tilde s}^\dagger(f) \ \mat{J}^{-1}(f) \ \vec{\tilde s}(f)} & f \neq 0 \\ 0  & f=0
    \end{cases},
\end{equation}
where the tilde denotes the Fourier transform and $\mat{J}(f)$ is the power spectral density matrix of the noise.

The photon energy which minimizes \cref{eq:chi2} can then be retrieved by finding the maximum of the convolution between the filter and the data and transforming from response amplitude to energy by inverting $A(E)$.
\begin{equation} \label{eq:e_hat}
    \hat E = A^{-1}\parens*{\max_t \braces*{\brackets*{\vec{h}^\transpose * \vec{d}](t)}}}
\end{equation}

\section{Coordinate Transformation} \label{sec:coordinates}
For phase responses greater than \SI{\sim 110}{\degree}, the coordinates defined by \cref{eq:Theta1,eq:D1} introduce an energy dependent response shape in violation of the assumptions used to derive \cref{eq:e_hat}. If we could use phase and dissipation coordinates that do not have these artificial effects, energy estimation with the filtering method may become more accurate. With this motivation, we introduce a new coordinate system. 

Since we expect $\sfrac{\delta f_r}{f_r}$ and $\delta Q_i^{-1}$ to be proportional to the photon energy, these variables are, perhaps, the most obvious choice for the new coordinates. However, the reactive current nonlinearity described by \cref{eq:x_nonlinear} requires a real-time estimate of the internal quality factor to solve for $\sfrac{\delta f_r}{f_r}$, amplifying the noise. Instead, we use $\delta x$ as the phase component, which is more linear with photon energy than $\varTheta_1$ and approximately equal to $\sfrac{-\delta f_r}{f_r}$ if no reactive nonlinearity exists. $\delta x$ and $\delta Q_i^{-1}$ can be solved for in terms of \sI\ and \sQ\ by using the adiabatic equivalence between the mixer output, \sZ, and $S_{21}$. 

\begin{subequations}
\begin{align}
    \varTheta_2(t) &\equiv \frac{-4 Q}{1 + 4 Q^2 x^2} \; \delta x(t) \notag \\
    &=\frac{-4 Q}{1 + 4 Q^2 x^2} \brackets*{\frac{\sQ(t) + 2 Q_c x_a \parens*{\sI(t) - 1}}{2 Q_c \abs*{1 - \sZ(t)}^2} - x} \label{eq:Theta2}\\
    D_2(t) & \equiv \frac{-2 Q}{1 + 4 Q^2 x^2} \; \delta Q_i^{-1}(t) \notag \\
    &= \frac{-2 Q}{1 + 4 Q^2 x^2} \brackets*{\frac{\sI(t) - \abs*{\sZ(t)}^2 + 2 Q_c x_a \sQ(t)}{Q_c \abs*{1 - \sZ(t)}^2} - Q_i^{-1}} \label{eq:D2}
\end{align}
\end{subequations}
The prefactor in both equations normalizes $\varTheta_2$ and $D_2$ to the polar coordinates in the small signal limit using \cref{eq:dTheta1,eq:dD1}. The normalization facilitates the comparison between the two transformations, shown in \cref{fig:coordinates}, by making them equal near the origin, but it is not critical to the implementation of this method. 

This new coordinate system compensates for the saturation in both the phase and dissipation signal, making a linear model for the photon response more valid. Additionally, for MKIDs where the majority of the noise power in the relevant bandwidth comes from two-level systems, the noise becomes stationary. The power spectral density of the noise, $\mat{J}(f)$, can then be used in place of the covariance matrix $\mat{C}(t, t'; \vec{\xi})$. In contrast, the amplifier noise becomes nonstationary. Care should be taken to properly model the noise if this component limits the resolving power.

\begin{figure}[t] \centering
    \includegraphics[width=0.8\linewidth]{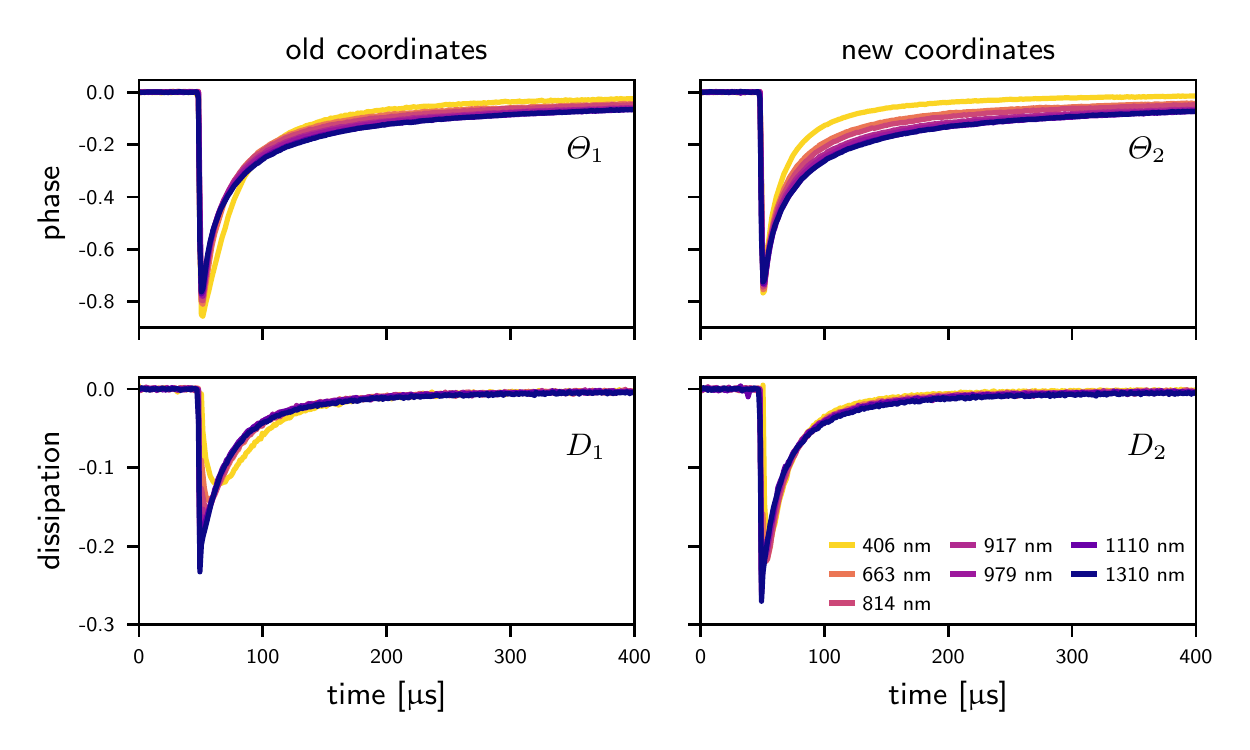}
    \caption{Shown are averaged photon responses of a hafnium optical to near-IR MKID for seven different laser energies. The functions are normalized to be compatible with \cref{eq:linear_model} as $\vec{s}(t)$ and should all lie on top of each other if the response shape is a constant function of energy. \textit{Left:} The old coordinates, $\varTheta_1$ and $D_1$, both show a strong energy dependence. \textit{Right:} The energy dependence is almost fully removed for $D_2$. $\varTheta_2$ still depends on energy but shows the expected trend of a faster decay for higher energy photons due to the higher quasiparticle density.}
    \label{fig:templates}
\end{figure}

\section{Application to Data} \label{sec:data}
To test the effectiveness of \cref{eq:Theta2,eq:D2}, we analyze data from a hafnium MKID array designed to be most sensitive to photons in the \SIrange{800}{1300}{nm} wavelength range. This device has been previously characterized in reference~\citenum{Zobrist2019a} using a high-electron-mobility transistor (HEMT) amplifier as the lowest temperature amplification stage. In order improve the data quality, we use the parametric amplifier readout discussed in reference~\citenum{Zobrist2019} which lowers the amplifier noise to near the quantum limit. 

In this configuration, the time domain noise---from either two-level systems or amplifiers---no longer dominates the uncertainty in the measured photon energy.~\cite{Zobrist2019} Instead, phonon losses to the substrate, a nonuniform detector response, and response shape nonlinearities limit the achievable resolving power for the detector. Therefore, no significant penalty is incurred by ignoring the nonstationary aspects of the noise and using \cref{eq:filter} to calculate the filter. If this were not the case, more care would need to be taken to estimate the full covariance matrix when constructing the filter.

The potential template functions for use in computing the filter are shown in \cref{fig:templates} for several photon energies in each coordinate system. If the linear photon response model from \cref{eq:linear_model} were fully satisfied, these templates would be identical for each energy. We see that the compression in the phase and dissipation coordinates is successfully corrected with $\varTheta_2$ and $D_2$, but some nonlinearities still exist. In particular, both the response ring-up and decay times are nonthermal which lead to an intrinsic energy dependence in the response shape. Additionally, for higher energy photons, the onset of the dissipation response becomes slightly delayed with respect to that of the phase response. Some improvement to the analysis may be achieved by properly accounting for these effects. However, since we are only evaluating the effectiveness of the new coordinate system with respect to the linear filtering method, we will not discuss them further and use the \SI{1110}{nm} template as $\vec{s}(t)$ for both coordinates and all photon energies. This choice is somewhat arbitrary and the results do not change significantly if any of the five lower energy functions are chosen. The two highest energy functions are avoided because they are much less representative of the larger dataset.

\begin{figure}[t] \centering
    \includegraphics[width=0.6\linewidth]{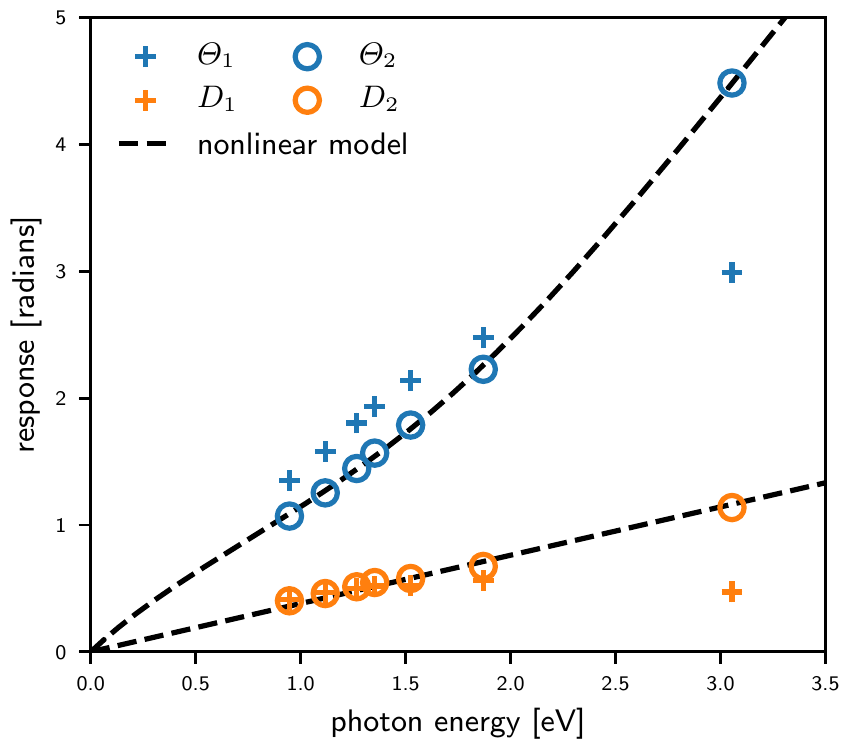}
    \caption{The laser calibration energies and detector responses used to construct the calibration function $A(E)$ (\cref{eq:linear_model}) are plotted for both coordinate systems.
    The responses in the $\varTheta_1$ and $D_1$ coordinates saturate at the higher energies while the $\varTheta_2$ and $D_2$ coordinates undo this effect. Any remaining nonlinear behavior in $\varTheta_2$ is well described by \cref{eq:x_nonlinear} and the fit parameters from \cref{eq:S21} (dashed black line).}
    \label{fig:calibration}
\end{figure}

The points used to construct the calibration function, $A(E)$, for each coordinate system are shown in \cref{fig:calibration}. As expected the new coordinates significantly improve the detector linearity. The dissipation response becomes completely linear, and the saturation in $\varTheta_1$ above \SI{1.6}{eV} is removed. However, the reactive current nonlinearity in the resonator causes $\varTheta_2$ to become nonlinear, resulting in small changes to the response shape across the measured energy range. Since there are already much larger response shape energy dependencies in the data, this effect does not influence the resolving power.

With both the template and calibration, we use \cref{eq:e_hat} to estimate the photon energies from the seven lasers. The resolving power is extracted by approximating the resulting distributions with a Gaussian kernel density estimate. The results are presented in \cref{tab:R} for four different coordinate combinations. 

For the five lowest energies using only phase data, $\varTheta_2$ provides no benefit over $\varTheta_1$ and actually performs slightly worse. The small amount of signal saturation in $\varTheta_1$ helps to counteract the energy dependence in the response shape. As expected, however, the removal of the response shape compression allows $\varTheta_2$ to outperform $\varTheta_1$ at the highest two energies. Moreover, using both the phase and dissipation coordinates generally improves the resolving power over just using the phase coordinate, but at the highest two energies the nonlinearity in $D_1$ is strong enough to reduce the resolving power. This effect is not seen with the $\varTheta_2$ and $D_2$ coordinates, and they produce the best results over the entire energy range.

\ctable[
caption={The resolving powers are given for a hafnium optical to near-IR MKID in different conditions. Seven different photon energies and four different coordinate combinations are presented. The new phase and dissipation coordinates, $\varTheta_2$ and $D_2$, outperform the other options.},
label={tab:R},
width=0.8\columnwidth,
doinside=\renewcommand*{\arraystretch}{1},
botcap
]{
l l C C C C
}{
}{
\toprule
\multicolumn{2}{c}{\multirow{2}*{Energy $\brackets*{\si{eV}}$}} & \multicolumn{4}{c}{Resolving Power $\brackets*{\sfrac{E}{\Delta E}}$}  \\
\cmidrule(l{1em}){3-6}
& & $\varTheta_1$ & $\varTheta_1$ \& $D_1$ & $\varTheta_2$ & $\varTheta_2$ \& $D_2$ \\
\cmidrule(r{.5em}){1-2} \cmidrule(l{1em}r{1em}){3-3} \cmidrule{4-4} \cmidrule(l{1em}r{1em}){5-5} \cmidrule{6-6}
\num{3.03}  & (\SI{406}{nm})  & \num{6.7} & \num{5.0} & \num{14}  & \num{17}  \\
\num{1.87}  & (\SI{663}{nm})  & \num{9.9} & \num{8.6} & \num{13}  & \num{13}  \\
\num{1.52}  & (\SI{814}{nm})  & \num{12}  & \num{13}  & \num{9.6} & \num{12}  \\
\num{1.35}  & (\SI{917}{nm})  & \num{11}  & \num{12}  & \num{9.3} & \num{12}  \\
\num{1.27}  & (\SI{979}{nm})  & \num{11}  & \num{12}  & \num{9.6} & \num{12}  \\
\num{1.12}  & (\SI{1110}{nm}) & \num{9.8} & \num{9.4} & \num{8.3} & \num{10}  \\
\num{0.946} & (\SI{1310}{nm}) & \num{8.4} & \num{8.4} & \num{8.1} & \num{9.6} \\
\bottomrule
}

\section{Conclusion} \label{sec:conclusion}
The traditionally used polar coordinates for analyzing MKID data have significant shortcomings for large signals in single photon detectors. The resulting signal compression is an artifact of the homodyne readout scheme and not of the underlying detector response. It is, therefore, possible to define a set of coordinates which removes this compression. This benefit is of primary importance to detectors whose data must be analyzed in real-time but may also be useful in combination with more complex algorithms by effectively reducing one source of nonlinearity.

We have shown that these new coordinates significantly extend the dynamic range of the current generation of optical to near-IR MKIDs when using the linear filtering method for extracting the photon energy. The resolving power at \SI{400}{nm} is more than doubled without needing to rely on more computationally expensive algorithms to account for the energy-dependent response shape. Further sources of response shape nonlinearity may also become more easily modeled in the new coordinate system as it is more clearly linked to the detector response.

Other types of single photon counting MKIDs could also benefit from this coordinate transformation. In particular, thermal kinetic inductance detectors do not have any of the residual response shape nonlinearities seen by optical to near-IR MKIDs and may see all of the response shape energy dependence removed by this method. Any resulting improvement to the resolution of these detectors, however, would depend on the response shape nonlinearity being the primary source of uncertainty in the measured photon energy.

\subsection*{Acknowledgments}
This research was carried out in part at the Jet Propulsion Laboratory and California Institute of Technology, under a contract with the National Aeronautics and Space Administration. N. Z. was supported throughout this work by a NASA Space Technology Research Fellowship. The MKID array used was developed under NASA grant NNX16AE98G.

\bibliography{references}   
\bibliographystyle{spiejour}   

\end{document}